\documentclass{cargese}\usepackage{graphicx}

\let\footnote\savefootnote
\let\footnotetext\savefootnotetext
\let\footnoterule\savefootnoterule
\normallatexbib
\newcommand{\hh}[1]{{[hep-th/{#1}]}}

\newcommand{\NPB}[1]{Nucl. Phys.~{\bf B#1}}
\newcommand{\PLB}[1]{Phys. Lett.~{\bf B#1}}

\newcommand{\CMP}[1]{Comm. Math. Phys.~{\bf #1}}

\newcommand{\MPL}[1]{Mod. Phys. Lett.~{\bf #1}}

\def\CA{{\cal A}}

\def\CE{{\cal E}}
\def\CF{{\cal F}}

\def\CM{{\cal M}}

\def\CS{{\cal S}}

\def\CW{{\cal W}}

\def\a{{\alpha}}
\def\b{{\beta}}

\def\f{{\phi}}	\def\varf{{\varphi}}
\def\g{{\gamma}}

\def\l{{\lambda}}
\def\m{{\mu}}
\def\n{{\nu}}

		 \def\P{{\Psi}}

\def\zb{{\zbeta}}

\def\be{\mu^z_\zb}
\def\beb{\mu^\zb_z}

\def\ab{{\bar a}}
\def\zb{{\bar z}}
\def\LB{{\bar{L}}}

\def\demi{{1\over 2}}

\def\gst{g_{{}_{\rm s}}}

\newcommand\beq{\begin{equation}}
\newcommand\eeq{\end{equation}}
\newcommand\bea{\begin{eqnarray}}
\newcommand\eea{\end{eqnarray}}

\renewcommand{\theequation}{\arabic{equation}}

\def\mytitle{{From Topological Field Theories
to Covariant Matrix Strings}}
\def\shorttitle{{}}

\newcommand\lpthe{{LPTHE, Universit\'es Paris VI \& VII,
 4 place Jussieu, 75252 Paris Cedex 05, France \\}}

\newcommand\sissa{{International School for Advanced Studies,
 2--4 via Beirut, 34014 Trieste, Italy \\}}

\newcommand\nikaf{{Lyman Laboratory for Physics, Harvard University, 
Cambridge MA 02138, USA \\}}

\begin{document}

\articletitle[]{{\mytitle}\footnote{Gong Show talk presented by C\'eline
Laroche at the  Carg\`ese'99 summer school ``Progress in String Theory and
M-theory",  May 24 - June 5, 1999.}}

%% optional, to supply a shorter version of the
%% title for the running head:

\chaptitlerunninghead{{\shorttitle}}

\author{Laurent Baulieu${}^{(1)}$,
C\'eline Laroche${}^{(1)(2)}$,
Nikita Nekrasov${}^{(1)(3)}$}

\affil{${}^{(1)}$ \lpthe  \\
${}^{(2)}$ \sissa \\
${}^{(3)}$ \nikaf}

\email{baulieu@lpthe.jussieu.fr,
laroche@sissa.it,
nikita@curie.harvard.edu \\}

\begin{abstract}
This paper is a shortened version of the previous work~\cite{bln}:
We propose a topological quantum field theory as a twisted candidate 
to formulate covariant matrix strings. The model relies on the octonionic 
or complexified instanton equations defined on an eight dimensional 
manifold with reduced holonomy. To allow untwisting of the model 
without producing an anomaly, we suggest  (partially twisted) $\CW$-gravity 
as an ``extended'' 2d-gravity sector.
\end{abstract}

The covariantization with respect to both worldsheet and spacetime 
symmetries
  of the orbifold theory of Dijkgraaf, Verlinde and Verlinde~\cite{dvv}, 
if it can be  achieved, should produce a covariant formulation of the 
second quantized free string
 theory. The statement that these  covariant matrix strings
  exist is quite
obvious in the commuting limit.
 Indeed,
consider the superconformal theory constructed as the sum of
$N$ copies of the standard Neveu--Schwarz--Ramond (NSR) superstring.  
It has
central charge
$c=N\times 0$ and involves the following sets of fields: bosons
$X^\m_i$, left and right moving fermions $\psi^\m_{i;L,R}$ and $2d$ 
gravity
ghosts
$(b_i, c_i),(\b_i,\g_i)$.  The indices $\m,
\n =1\cdots 10$ are acted on by the ten-dimensional Lorentz group,  
and the
index
$i=1\cdots N$ by the  ${\CS_N}$ symmetry group. This theory, whose  
action is
simply the sum of
$N$ NSR superstrings' actions,
 can be ``orbifolded'' with respect to the
${\CS_N}$ symmetry (add twisted sectors ---the long strings---  and  
project onto
the invariant states). Orbifolding preserves both the manifest  
ten-dimensional
and (less manifest) two-dimensional covariance, and the  theory it leads 
to has all the nice
features of a covariant string theory. It exhibits a $\CW$-like
chiral algebra,
generated by the stress energy tensor $T=\sum_i T_i$ and  its  
higher spin
analogues $\CW_{l=1\cdots N}=\sum_i T_i^l$.
Furthermore, one can show that on the cylinder, the theory is  
equivalent to
the Green--Schwarz theory of~\cite{dvv}.

The tricky question is of course that of going  away from the  
$\gst=0$ limit.
In~\cite{bln}, we make a step and
 address this issue from the ``twisted'' point of view. Namely, we  
propose a
topological quantum field theory (TQFT) as a twisted candidate,
possibly invariant under $SO(8)\times SO(2)$.
The TQFT is based
on the generalized
octonionic or complexified instanton equations defined in eight
dimensions for a
$U(N)$ Yang--Mills field~\cite{bks&acha} {\sl and } superpartners,
 living on an
eight dimensional manifold~$\CM_8$ with special holonomy.

The octonionic instantons arise from the
generalized self-duality  equations that occur when $\CM_8$ has  
$Spin(7)$
holonomy:
\bea\label{oct}
{\CF_{i8} \pm \demi c_{ijk} \CF_{jk}=0, \qquad {\rm for} \quad
i,j,\cdots = 1, \ldots, 7}
\eea
Where $\CF$ denotes the curvature of the gauge field $\CA$, and the
$c_{ijk}$ are the structure constants for octonions.\\
In the case of $\CM_8$ being a
Calabi--Yau fourfold, it is the $SU(4)$ instantons one deals with,  
obtained
from
\bea
\label{cpx}
\begin{array}{c}
\CF_{z^A \zb^A} =0,  \\
\CF_{z^A z^B} \pm
\epsilon_{{}_{ABCD}} \CF_{\zb^C \zb^D} =0,
\end{array}
\quad {\rm for}\quad A,B, \cdots=1, \ldots,4
\eea

Such equations can be used as gauge functions to define
topological field theories~\cite{selfd} in which the topological
symmetry  can
possibly be untwisted into supersymmetry, in the same spirit
as~\cite{fourd}.  The theories
of~\cite{bks&acha} have
been  constructed in this manner and, after reduction to
two dimensions, provide a
twisted version of the light-cone matrix strings of~\cite{dvv}.
Their topological multiplet can be untwisted into the degrees of  
freedom of
the light-cone  matrix string under
the following correspondence, inspired of~\cite{bgr}
(according to the splitting $\CM_{8} = \CM_{6}
\times \Sigma$ where
$\Sigma$ is  two--dimensional
and $\CM_6$ is taken to be small, the gauge
field
$\CA_{z^{A=1\cdots4}}$ splits as $\f_a \sim \CA_{z^{a=1\cdots 3}}$
and  $A_z \sim
\CA_{z^4}$):
\bea
\begin{array}{rll}
\f^a,\, \bar{\f}^a,\, \varf,\, \bar\varf\,
 \ & \to &\ X^{m=1\cdots 8},\\
 \P^a,\, \chi^\ab,\, \eta + \chi,\, \P_z \, &\to & \
\psi_{{}_L}^m, \\
 \P^\ab,\, \chi^a,\, \eta - \chi,\, \P_\zb\, & \to & \
\psi_{{}_R}^m .
\end{array}
\eea
Here, for each classical field ($A_{z, \zb}, \f, \bar\f $), the  
$\P$'s are the
topological ghosts and the $\chi$'s are the antighosts.  The  
Yang--Mills field
also involves  ghosts for ghosts (the
$\varf $'s), responsible for the gauge symmetry, and the associated  
fermionic
Lagrange multiplier $\eta$.
All fields are  $U(N)$ valued; the $\f^a$'s are
complex and we note $\f^{\bar a} \sim \bar \f ^a \sim {\f^a}^\dag$.

To go beyond the light-cone 
version, we make two distinct
steps.  The
first one, concerning ten-dimensional spacetime covariance, is to  
enhance
the field
content to that of the whole eight-dimensional Yang--Mills  
supermultiplet,
to end  with the degrees of freedom of a ten-dimensional target space.
Hence, instead of constructing the TQFT for the sole $U(N)$ gauge field 
$\CA_\m$, we
include its eight dimensional superpartners
$\l^\a, \f^\pm$.% (together with their  BRST multiplets). 
The  scalars $\f^\pm$ will provide  the longitudinal coordinates, whereas
the fermions $\l^\a$  (and topological partners) 
open the gate to a heterotic  sector. 

This in turn {\sl naturally} leads to the second step, that of
two dimensional covariance. Indeed, one needs to gauge fix the  
topological
symmetry of
the fermions and scalars, and this can only be done if we dimensionally 
reduce down to
two dimensions. However, simplistic reduction turns our gauge
functions~(\ref{oct})(\ref{cpx}) into
a set of inconsistent  equations.
Two-dimensional covariance implies the  coupling   to a
topological
$2d$ gravity system.
The metric  is introduced in the Beltrami
parametrization (it allows for manifest left-right
factorization~\cite{beltrami}): $
ds^2 = e_z^+e_\zb^- (dz +\be d\zb)(d\zb +\beb dz)$ and
the symmetries are used to set the background values of
$ e_z^+, e_\zb^- $ to 1.
To make our gauge functions covariant, we   introduce  additional   
fields, of
the Wess--Zumino type. Denoted as $L, \LB$, their  conformal  
weight is
fixed by
the requirement that $e^L$ and $e^\LB$ have weight $(1,0)$ and $(0,1)$ 
respectively. Making use of these, we can  define our two dimensional
covariant gauge functions, based on~(\ref{cpx}), as follows:
\bea\label{coveqs}
\begin{array}{cll}
 \CE &\sim & F_{z\zb} + e^{L+\LB} \sum{ {}_{{}_a} } [ \f^{a},
\f^{\bar a} ]\\
 \CE^{a}_{\zb} &\sim& D_{\zb} \f^{a} + \demi e^{\LB }
\varepsilon^{abc} [\f^{\bar b}, \f^{\bar c}]\\
\CE^{\bar a}_{z}
&\sim&  D_z \f^{\bar a} + \demi e^{L}
\varepsilon^{abc} [\f^{b}, \f^{c}]
\end{array}
\eea
with derivatives  now   covariant with respect to both the gauge 
invariance and the
$2d$ gravity (they involve $A_z, A_\zb$, the Beltrami's 
and the Christoffel's).
As for the superpartners, we choose to impose the right-moving condition 
on the fermions  and holomorphicity on the scalars.
%\\
We have imposed two-dimensional covariance; we deal with the  degrees of
freedom of a ten dimensional target space  theory via
$ X^{M=1\cdots 10}
\leftarrow \{  X^{m},
\f^\pm \}$ for bosons, and   $\psi^M_{{}_L}
\leftarrow
\{\psi^m_{{}_L}, \P^+, \chi^- \}$, $\psi^M_{{}_R}
\leftarrow
\{\psi^m_{{}_R}, \P^-, \chi^+ \}$ for~fermions. 
%left and right movers, respectively.

 The key point is then to check whether we are allowed to map our
fields to  physical ones as above.
Indeed, it is crucial that we are able to untwist our (topological  
and thus
anomaly free) theory into a
supersymmetric one without producing an anomaly:
Can we change the spins of our fields to the physical
ones without spoiling the vanishing of the central~charge?
%\\
In the usual $U(1)$ superstrings, the light-cone theory is a gauge fixed
version where the longitudinal components are  canceled by the
$2d$ gravity
ghost systems arising from two dimensional reparametrizations.
For non-commuting
strings, it is not clear how one can make sense of
``matrix-valued'' $2d$-diffeomorphisms, nor what kind of ``extended''
 ghost system would be
capable of compensating for the longitudinal $U(N)$ degrees of freedom 
(namely $2(N^2-1)$ non-physical bosons, and  correspondingly for the
fermions).
\\
We refer the reader to~\cite{bln}, where we propose evidence that the
appropriate
ghost systems might  arise  from the $\CW_{N+1}$ of  $\CW_\infty$
partially twisted gravity.
Indeed, the TQFT based on the Lie algebra $\CW_\infty$ has an infinite
series of 
ghosts that could be partially untwisted to produce the
set of gravitational ghost systems to compensate for the matrix string's
matter anomaly.   
Different twists and  gauge fixings
 might allow  to get  either a covariant NSR  theory
{\sl or}  heterotic matrix strings (via more complicated compensations  
involving the fermions $\l^\a$, 
in the spirit of~\cite{bgr}).

\begin{acknowledgments}
It is a pleasure for C.L. to thank all the
organizers of Carg\`ese'99 for a very pleasant and stimulating
school, a refreshing gong show and generous support. C.L. also 
acknowledges EC TMR contract n.960090. The research of 
N.N. is partly supported by RFFI grant 98-01-00327, and 96-15-96455 for
scientific schools.
\end{acknowledgments}

%% appendix optional
%\chapappendix{This is the Appendix Title}
%This is an appendix with a title.

%\chapappendix{}
%This is an appendix without a title.

%%%%%%% BIBLIO

\end{document}